\documentclass{sig-alternate}
\usepackage{amsfonts}
\usepackage{listings}
\usepackage{amsmath}
\usepackage{amssymb}
\usepackage{graphicx}
\newcommand{\ignore}[1]{}

\begin{document}

\title{MapReduce is Good Enough? If All You Have is a\\ Hammer, Throw Away Everything That's Not a Nail!}

\author{\alignauthor Jimmy Lin \\
\affaddr{University of Maryland}\\
\affaddr{jimmylin@umd.edu}\\
\vspace{0.2cm}
Version of \today}

\date{}

\maketitle

\begin{abstract}
Hadoop is currently the large-scale data analysis ``hammer'' of
choice, but there exist classes of algorithms that aren't ``nails'',
in the sense that they are not particularly amenable to the MapReduce
programming model.  To address this, researchers have proposed
MapReduce extensions or alternative programming models in which these
algorithms can be elegantly expressed. This essay espouses a very
different position:\ that MapReduce is ``good enough'', and that
instead of trying to invent screwdrivers, we should simply get rid of
everything that's not a nail. To be more specific, much discussion in
the literature surrounds the fact that iterative algorithms are a poor
fit for MapReduce:\ the simple solution is to find alternative
non-iterative algorithms that solve the same problem.  This essay
captures my personal experiences as an academic researcher as well as
a software engineer in a ``real-world'' production analytics
environment.  From this combined perspective I reflect on the current
state and future of ``big data'' research.\\

\noindent {\bf Author's note:} I wrote this essay specifically to be
controversial. The views expressed herein are more extreme than what I
believe personally, written primarily for the purposes of provoking
discussion. If after reading this essay you have a strong
reaction, then I've accomplished my goal :)

\end{abstract}

\numberofauthors{1}
\additionalauthors{}

\section{Introduction}

MapReduce~\cite{Dean_Ghemawat_OSDI2004} has become an ubiquitous
framework for large-scale data processing. The Hadoop open-source
implementation enjoys widespread adoption in organizations ranging
from two-person startups to Fortune 500 companies. It lies at the core
of an emerging stack for data analytics, with support from industry
heavyweights such as IBM, Microsoft, and Oracle. Among the advantages
of MapReduce are the ability to horizontally scale to petabytes of
data on thousands of commodity servers, easy-to-understand programming
semantics, and a high degree of fault tolerance.

MapReduce, of course, is not a silver bullet, and there has been much
work probing its limitations, both from a theoretical
perspective~\cite{Karloff_etal_2010,Afrati_etal_2012} and empirically
by exploring classes of algorithms that cannot be efficiently
implemented with
it~\cite{Ekanayake_Fox_2009,Bhatotia_etal_2011,BuYingyi_etal_VLDB2010,ZhangYanfeng_etal_SoCC2011}. Many
of these empirical studies take the following form:\ they present a
class of algorithms for which the na\"ive Hadoop solution performs
poorly, expose it as a fundamental limitation of the MapReduce
programming model,\footnote{\small Note that in this paper I attempt
  to be precise when referring to MapReduce, the programming model,
  and Hadoop, the popular open-source implementations.} and propose an
extension or alternative that addresses the limitation. The algorithms
are expressed in this new framework, and, of course, experiments show
substantial (an order of magnitude!)\ performance improvements over
Hadoop.

This essay espouses a very different position, that MapReduce is
``good enough'' (even if the current Hadoop implementation could be
vastly improved).  While it is true that a large class of algorithms
are not amenable to MapReduce implementations, there exist alternative
solutions to the same underlying problems that {\it can} be easily
implemented in MapReduce. Staying in its confines allows more
tightly-integrated, robust, end-to-end solutions to heterogeneous
large-data challenges.

To apply a metaphor:\ Hadoop is currently the large-scale data
processing hammer of choice. We've discovered that, in addition to
nails, there are actually screws---and it doesn't seem like hammering
screws is a good idea. So instead of trying to invent a screwdriver,
let's just get rid of the screws.  If there are only nails, then our
MapReduce hammer will work just fine. To be specific, much discussion
in the literature surrounds the fact that iterative algorithms are not
amenable to MapReduce:\ the (simple) solution, I suggest, is to avoid
iterative algorithms!

I will attempt to support this somewhat radical thesis by exploring
three large classes of problems which serve as the poster children for
MapReduce-bashing:\ iterative graph algorithms (e.g., PageRank),
gradient descent (e.g., for training logistic regression classifiers),
and expectation maximization (e.g., for training hidden Markov models,
$k$-means).  I begin with vague and imprecise notions of what
``amenable'' and ``good enough'' mean, but propose a concrete
objective with which to evaluate competing solutions later.

This essay captures my personal experiences as an academic researcher
as well as a software engineer in a production analytics
environment. As an academic, I've been fortunate enough to collaborate
with many wonderful colleagues and students on ``big data'' since
2007, primarily using Hadoop to scale a variety of text- and
graph-processing algorithms (e.g., information retrieval, statistical
machine translation, DNA sequence assembly). Recently, I've just
returned from spending an extended two-year sabbatical at Twitter ``in the
trenches'' as a software engineer wrestling with various ``big data''
problems and trying to build scalable production solutions.

In earnest, I quip ``throw away everything not a nail''
tongue-in-cheek to make a point. More constructively, I suggest a
two-pronged approach to the development of ``big data'' systems and
frameworks.  Taking the metaphor a bit further (and at the expense of
overextending it):\ On the one hand, we should perfect the hammer we
already have by improving its weight balance, making a better grip,
etc. On the other hand, we should be developing jackhammers---entirely
new ``game changers'' that can do things MapReduce and Hadoop
fundamentally cannot do. In my opinion, it makes less sense to work on
solving classes of problems for which Hadoop is already ``good
enough''.

\section{Iterative Graph Algorithms}

Everyone's favorite example to illustrate the limitations of MapReduce
is PageRank (or more generally, iterative graph algorithms). Let's
assume a standard definition of a directed graph $G=(V,E)$ consisting
of vertices $V$ and directed edges $E$, with $S(v_i) = \{v_j |
(v_i,v_j) \in E \}$ and $P(v_i) = \{v_j | (v_j,v_i) \in E \}$
consisting of the set of all successors and predecessors of vertex
$v_i$ (outgoing and incoming edges, respectively).
PageRank~\cite{Page_etal_1999} is defined as the stationary
distribution over vertices by a random walk over the graph.  That is,
for each vertex $v_i$ in the graph, PageRank computes the value
$\textsc{Pr}(v_i)$ representing the likelihood that a random walk will
arrive at vertex $v_i$. This value is primarily induced from the graph
topology, but the computation also includes a damping factor $d$,
which allows for random jumps to any other vertex in the graph. For
non-trivial graphs, PageRank is generally computed iteratively over
multiple timesteps $t$ using the power method:
\begin{equation} \label{pagerank}
\textsc{Pr}(v_i;t) = \left\{
\begin{array}{l l}
1/|V| & \mbox{if $t = 0$} \\
\frac{1-d}{|V|} + d \sum_{v_j \in P(v_i)}{\frac{\textsc{Pr}(v_j;t-1)}{|S(v_j)|}} & 
\mbox{if $t > 0$} \\
\end{array}
\right.
\end{equation}
\noindent The algorithm iterates until either a user defined maximum
number of iterations has completed, or the values sufficiently converge.
One common convergence criterion is:
\begin{equation} \label{pagerankstop}
\sum{|\textsc{Pr}(v_i;t)-\textsc{Pr}(v_i;t-1)|} < \epsilon
\end{equation}
The standard MapReduce implementation of PageRank is well known and is
described in many places (see, for example,~\cite{Lin_Dyer_book}). The
graph is serialized as adjacency lists for each vertex, along with the
current PageRank value. Mappers process all the vertices in
parallel:\ for each vertex on the adjacency list, the mapper emits an
intermediate key-value pair with the destination vertex as the key and
the partial PageRank contribution as the value (i.e., each vertex
distributes its present PageRank value evenly to its successors). The
shuffle stage performs a large ``group by'', gathering all key-value
pairs with the same destination vertex, and each reducer sums up the
partial PageRank contributions.

Each iteration of PageRank corresponds to a MapReduce
job.\footnote{\small This glosses over the treatment of the random
  jump factor, which is not important for the purposes here, but
  see~\cite{Lin_Dyer_book}.}  Typically, running PageRank to
convergence requires dozens of iterations. This is usually handled by
a control program that sets up the MapReduce job, waits for it to
complete, and then checks for convergence by reading in the updated
PageRank vector and comparing it with the previous. This cycle repeats
until convergence.  Note that the basic structure of this algorithm
can be applied to a large class of ``message-passing'' graph
algorithms~\cite{Lin_Schatz_2010,Malewicz_etal_SIGMOD2010} (e.g.,
breadth-first search follows exactly the same form).

There is one critical detail necessary for the above approach to
work:\ the mapper must also emit the adjacency list with the vertex id
as the key.  This passes the graph structure to the reduce phase,
where it is reunited (i.e., joined) with the updated PageRank
values. Without this step, there would be no way to perform multiple
iterations.

There are many shortcoming with this algorithm:

\begin{list}{\labelitemi}{\leftmargin=1em}
\setlength{\itemsep}{-2pt}

\item MapReduce jobs have high startup costs (in Hadoop,
  can be tens of seconds on a large cluster under load). This places a
  lower bound on iteration time.

\item Scale-free graphs, whose edge distributions follow power laws,
  often create stragglers in the reduce phase. The highly uneven
  distribution of incoming edges to vertices produces significantly
  more work for some reduce tasks (take, for example, the
  reducer assigned to sum up the incoming PageRank contributions to
  {\tt \small google.com} in the webgraph). Note that since these
  stragglers are caused by data skew, speculative
  execution~\cite{Dean_Ghemawat_OSDI2004} cannot solve the problem.
  Combiners and other local aggregation techniques alleviate but do
  not fully solve this problem.

\item At each iteration, the algorithm must shuffle the graph
  structure (i.e., adjacency lists) from the mappers to the
  reducers. Since in most cases the graph structure is static, this
  represents wasted effort (sorting, network traffic, etc.).

\item The PageRank vector is serialized to HDFS, along with the graph
  structure, at each iteration. This provides excellent fault
  tolerance, but at the cost of performance.

\end{list}

\noindent To cope with these shortcomings, a number of extensions to
MapReduce or alternative programming models have been
proposed. Pregel~\cite{Malewicz_etal_SIGMOD2010} implements the Bulk
Synchronous Parallel model~\cite{Valiant_CACM1990}:\ computations are
``vertex-centric'' and algorithms proceed in supersteps with
synchronization barriers between each. In the implementation, all
state, including the graph structure, is retained in memory (with
periodic checkpointing). HaLoop~\cite{BuYingyi_etal_VLDB2010} is an
extension of Hadoop that provides support for iterative algorithms by
scheduling tasks across iterations in a manner that exploits data
locality and by adding various caching mechanisms. In
Twister~\cite{Ekanayake_Fox_2009}, another extension of Hadoop
designed for iteration, intermediate data are retained in memory if
possible, thus greatly reducing iteration
overhead. PrIter~\cite{ZhangYanfeng_etal_SoCC2011}, in contrast, takes
a slightly different approach to speeding up iterative
computation:\ it {\it prioritizes} those computations that are likely
to lead to convergence.

All the frameworks discussed above share in supporting iterative
constructs, and thus elegantly solve one or more of the shortcomings
of MapReduce discussed above. However, they all have one
drawback:\ they're not Hadoop! The reality is that the Hadoop-based
stack (e.g., Pig, Hive, etc.)\ has already gained critical mass as the
data processing framework of choice, and there are non-trivial costs
for adopting a {\it separate} framework just for graph processing or
iterative algorithms. More on this point in
Section~\ref{section:good-enough}. For now, consider three
additional factors:

First, without completely abandoning MapReduce, there are a few simple
``tweaks'' that one can adopt to speed up iterative graph
algorithms. For example, the Schimmy pattern~\cite{Lin_Schatz_2010}
avoids the need to shuffle the graph structure by consistent
partitioning and performing a parallel merge join between the graph
structure and incoming graph messages in the reduce phase. The authors
also show that great gains can be obtained by simple partitioning
schemes that increase opportunities for partial aggregation.

Second, some of the shortcomings of PageRank in MapReduce are not as
severe as the literature would suggest. In a real-world context,
PageRank (or any iterative graph algorithm) is almost never computed
from scratch, i.e., initialized with a uniform distribution over all
vertices and run until convergence. Typically, the previously-computed
PageRank vector is supplied as a starting point on an updated
graph. For example, in the webgraph context, the hyperlink structure
is updated periodically from freshly-crawled pages and the task is to
compute updated PageRank values. It makes little sense to
re-initialize the PageRank vector and ``start over''. Initializing the
algorithm with the previously-computed values significantly reduces
the number of iterations required to converge. Thus, the iteration
penalties associated with MapReduce become much more tolerable.

Third, the existence of graph streaming algorithms for computing
PageRank~\cite{Sarma_etal_PODS2008} suggests that there may be
non-iterative solutions (or at least approximations thereof) to a
large number of iterative graph algorithms. This, combined with a good
starting distribution (previous point), suggests that we can compute
solutions efficiently, even within the confines of MapReduce.

Given these observations, perhaps we might consider MapReduce to be
``good enough'' for iterative graph algorithms? But what exactly does
``good enough'' mean? Let's return to this point in
Section~\ref{section:good-enough}.

\section{Gradient Descent}

Gradient descent (and related quasi-Newton) methods for machine
learning represent a second large class of problems that are poorly
suited for MapReduce. To explain, let's consider a specific type of
machine learning problem, supervised classification. We define $X$ to be
an input space and $Y$ to be an output space. Given a set of training
samples $D=\{(\mathrm{x}_{i},y_{i})\}_{i=1}^{n}$ from the space $X
\times Y$, the task is to induce a function $f : X \rightarrow Y$ that
best explains the training data. The notion of ``best'' is usually
captured in terms of minimizing ``loss'', via a function $\ell$ that
quantifies the discrepancy between the functional prediction
$f(x_{i})$ and the actual output $y_{i}$, for example, minimizing the
quantity:
\begin{equation}
\frac{1}{n} \sum_{i=0}^{n} \ell(f(x_{i}), y_{i})
\end{equation}
\noindent which is known as the {\it empirical risk}. Usually, we
consider a family of functions $\mathcal{F}$ (i.e., the hypothesis
space) that is parameterized by the vector $\theta$, from which we
select:
\begin{equation}
\arg \min_{\mathrm{\theta}} \frac{1}{n} \sum_{i=0}^{n} \ell(f(x_{i}; \theta), y_{i})
\label{loss}
\end{equation}
\noindent That is, we learn the parameters of a particular model. In
other words, machine learning is cast as a functional optimization
problem, often solved with gradient descent.

Rewriting Equation~(\ref{loss}) as $\arg \min_{\mathrm{w}} L(\theta)$
simplifies our notation.  The gradient of $L$, denote $\nabla L$, is
defined as follows:
\begin{equation}
\nabla L(\theta) = \left[ \frac{\partial L(\theta)}{\partial w_0} , 
\frac{\partial L(\theta)}{\partial w_1} , \ldots
\frac{\partial L(\theta)}{\partial w_d}
\right]
\end{equation}
The gradient defines a vector field pointing to the direction in which
$L$ is increasing the fastest and whose magnitude indicates the rate
of increase. Thus, if we ``take a step'' in the direction opposite to
the gradient from an arbitrary point $\mathrm{a}$,
$\mathrm{b} = \mathrm{a} - \gamma \nabla L(\textrm{a})$,
\noindent then $L(\mathrm{a}) \ge L(\mathrm{b})$, provided that
$\gamma$ (known as the step size) is a small value greater than zero.

If we start with an initial guess of $\theta^{(0)}$ and repeat the
above process, we arrive at gradient descent. More formally, let us
consider the sequence $\theta^{(0)}, \theta^{(1)}, \theta^{(2)}
\ldots$, defined with the following update rule:
\begin{equation}
\theta^{(t+1)} \leftarrow \theta^{(t)} - \gamma^{(t)} \nabla L(\theta^{(t)})
\end{equation}
\noindent We have:
\begin{equation}
L(\theta^{(0)}) \ge L(\theta^{(1)}) \ge L(\theta^{(2)}) \ldots
\end{equation}
\noindent where the sequence converges to the desired local
minimum. If the loss function is convex and $\gamma$ is selected
carefully (which can vary per iteration), we are guaranteed to
converge to a global minimum.

Based on the observation that our loss function decomposes linearly,
and therefore the gradient as well, the MapReduce implementation of
gradient descent is fairly straightforward. We process each training
example in parallel and compute its partial contribution to the
gradient, which is emitted as an intermediate key-value pair and
shuffled to a single reducer. The reducer sums up all partial gradient
contributions and updates the model parameters.  Thus, each iteration
of gradient descent corresponds to a MapReduce job. Two more items are
needed to make this work:

\begin{list}{\labelitemi}{\leftmargin=1em}
\setlength{\itemsep}{-2pt}

\item Complete classifier training requires many MapReduce jobs to be
  chained in a sequence (hundreds, even thousands, depending on the
  complexity of the problem). Just as in the PageRank case, this is
  usually handled by a driver program that sets up a MapReduce job,
  waits for it to complete, and then checks for convergence, repeating
  as long as necessary.

\item Since mappers compute partial gradients with respect to the
  training data, they require access to the current model
  parameters. Typically, the parameters are loaded in as ``side data''
  in each mapper (in Hadoop, either directly from HDFS or from the
  distributed cache). However, at the end of each iteration the
  parameters are updated, so it is important that the updated model is
  passed to the mappers at the next iteration.

\end{list}

\noindent Any number of fairly standard optimizations can be applied
to increase the efficiency of this implementation, for example,
combiners to perform partial aggregation or the in-mapper combining
pattern~\cite{Lin_Dyer_book}. As an alternative to performing gradient
descent in the reducer, we can substitute a quasi-Newton method such
as L-BFGS~\cite{LiuDong_etal_1989} (which is more expensive, but
converges in few iterations).  However, there are still a number of
drawbacks:

\begin{list}{\labelitemi}{\leftmargin=1em}
\setlength{\itemsep}{-2pt}

\item As with PageRank, Hadoop jobs have high startup costs.

\item Since the reducer must wait for all mappers to finish (i.e., all
  contributions to the gradient to arrive), the speed of each
  iteration is bound by the {\it slowest} mapper, and hence sensitive
  to stragglers. This is similar to the PageRank case, except in the
  map phase.

\item The combination of stragglers and using only a single reducer
  potentially causes poor cluster utilization. Of course, the cluster
  could be running other jobs, so from a throughput perspective, this
  is only a minor concern.

\end{list}

\noindent The shortcomings of gradient descent implementations in
MapReduce have prompted researchers to explore alternative
architectures and execution models that address these issues. All
the systems discussed previously in the context of PageRank are
certainly relevant, but we point out two more
alternatives. Spark~\cite{Zaharia_etal_NSDI2012} introduces the Resilient
Distributed Datasets (RDD) abstraction, which provide a restricted
form of shared memory based on coarse-grained transformations rather
than fine-grained updates to shared state. RDDs can either be cached
in memory or materialized from durable storage when needed (based on
lineage, which is the sequence of transformations applied to the
data).  Classifier training is one of the demo applications in
Spark. Another approach with similar goals is taken by Bu et
al.~\cite{BuYingyi_etal_2012}, who translate iterative
MapReduce and Pregel-style programs into recursive queries in
Datalog. By taking this approach, database query optimization
techniques can be used to identify efficient execution
plans. These plans are then executed on the Hyracks data-parallel
processing engine~\cite{Borkar_etal_ICDE2011}.

In contrast to these proposed solutions, consider an alternative
approach. Since the bottleneck in gradient descent is the iteration,
let's simply get rid of it! Instead of running {\it batch} gradient
descent to train classifiers, let us adopt {\it stochastic}
gradient descent, which is an {\it online} technique. The simple idea
is that instead of updating the model parameters after only
considering {\it every} training example, let us update the model
after {\it each} training example (i.e., compute the gradient with
respect to each example).

Online learning techniques have received renewed interest in the
context of big data since they operate in a streaming fashion and are
very
fast~\cite{Bottou_LeCun_NIPS2003,Shalev-Shwartz_etal_ICML2007,Langford_etal_JMLR2009,Bottou_2010}. In
practice, classifiers trained using online gradient descent achieve
accuracy comparable to classifiers trained using traditional batch
learning techniques, but are an order of magnitude (or more) faster to
train~\cite{Bottou_2010}.

Stochastic gradient descent addresses the iteration problem, but does
not solve the single reducer problem. For that, ensemble methods come
to the rescue~\cite{Dietterich97,Kuncheva_2004}. Instead of training a
single classifier, let us train an {\it ensemble} of classifiers and
combine predictions from each (e.g., simple majority voting, weighted
interpolation, etc.). The simplest way of building
ensembles---training each classifier on a partition of the training
examples---is both embarrassingly parallel and surprisingly effective
in practice~\cite{Mann_etal_NIPS2009,McDonald_etal_HLT-NAACL2010}.

Combining online learning with ensembles addresses the shortcomings of
gradient descent in MapReduce. As a case study, this is how Twitter
integrates machine learning into Pig in a scalable
fashion~\cite{Lin_Kolcz_SIGMOD2012}:\ folding the online learning
inside storage functions and building ensembles by controlling data
partitioning. To reiterate the argument:\ if MapReduce is not amenable
to a particular class of algorithms, let's simply find a different
class of algorithms that will solve the same problem and {\it is}
amenable to MapReduce.

\section{Expectation Maximization}

A third class of algorithms not amenable to MapReduce is expectation
maximization (EM)~\cite{Dempster_Laird_Rubin_1977} and EM-like
algorithms. Since EM is related to gradient descent (both are
first-order optimization techniques) and many of my arguments are
quite similar, the discussion in this section will be more
superficial.

EM is an iterative algorithm that finds a successive series of
parameter estimates $\theta^{(0)}$, $\theta^{(1)}$, $\ldots$ that
improve the marginal likelihood of the training data, used in cases
where there is incomplete (or unobservable) data. The algorithm starts
with some initial set of parameters $\theta^{(0)}$ and then updates
them using two steps:\ expectation (E-step), which computes the
posterior distribution over the latent variables given the observable
data and a set of parameters $\theta^{(i)}$, and maximization
(M-step), which computes new parameters $\theta^{(i+1)}$ maximizing
the expected log likelihood of the joint distribution with respect to
the distribution computed in the E-step. The process then repeats with
these new parameters. The algorithm terminates when the likelihood
remains unchanged.

Similar to iterative graph algorithms and gradient descent, each EM
iteration is typically implemented as a Hadoop job, with a driver to set up
the iterations and check for convergence. In broad strokes, the E-step
is performed in the mappers and the M-step is performed in the
reducers. This setup has all the shortcomings discussed before, and EM
and EM-like algorithms can be much more elegantly implemented in
alternative frameworks that better support iteration (e.g., those
presented above).

Let's more carefully consider terms that I've been using quite
vaguely:\ What does it mean for an algorithm to be {\it amenable} to
MapReduce? What does it mean for MapReduce to be ``good enough''? And
the point of comparison? Here are two case studies that build up to my
point:\

Dyer et al.~\cite{Dyer_etal_2008} applied
MapReduce to training translation models for a statistical machine
translation system---specifically, the word-alignment component that
uses hidden Markov models (HMMs) to discover word
correspondences across bilingual corpora~\cite{Tiedemann_2011}. The point of
comparison was GIZA++,\footnote{\small code.google.com/p/giza-pp/}
a widely-adopted in-memory, single-threaded implementation
(the {\it de facto} standard used by
researchers at the time the work was performed, and still commonly
used today).  The authors built a Hadoop-based implementation of the
HMM word-alignment algorithm, which demonstrated linear scalability
compared to GIZA++, reducing per-iteration training time from hours to
minutes. The implementation exhibited all the limitations associated
with EM algorithms (high job startup costs, awkward passing of model
parameters from one iteration to the next, etc.), yet compared to the
previous single-threaded approach, MapReduce represented a step
forward.\footnote{\small HMM training is relatively expensive
  computationally, so job startup costs are less of a
  concern. Furthermore, these algorithms typically run for less than a
  dozen iterations.} Here is the key point:\ whether an algorithm is
``amenable'' to MapReduce is a relative judgment that is only
meaningful in the context of an alternative. Compared to GIZA++, the
Hadoop implementation represented an advance.  However, this
is not inconsistent with the claim that EM algorithms could be more
elegantly implemented in an alternate model that better supports
iteration (e.g., any of the work discussed above).

The second example is the venerable Lloyd's method for $k$-means
clustering, which can be understood in terms of EM (not exactly EM,
but can be characterized as EM-like). A MapReduce implementation of
$k$-means shares many of the limitations discussed thus far. It is
true that the algorithm can be expressed in a simpler way using a
programming model with iterative constructs and executed more
efficiently with better iteration support (and indeed, many of the
papers discussed above use $k$-means as a
demo application). However, even within the confines of
MapReduce, there has been a lot of work on optimizing
clustering algorithms (e.g.,~\cite{Cordeiro_etal_SIGKDD2011,Ene_etal_SIGKDD2011}).
It is not entirely clear how these improvements
would stack up against using an entirely different framework. Here,
is MapReduce ``good enough''?

These two case studies provide the segue to my attempt at more clearly
defining what it means for MapReduce to be ``good enough'', and a
clear objective for deciding between competing solutions.

\section{What's ``Good Enough''?}
\label{section:good-enough}

I propose a pragmatic, operational, engineering-driven criterion for
deciding between alternative solutions to large-data problems. First,
though, my assumptions:

\begin{list}{\labelitemi}{\leftmargin=1em}
\setlength{\itemsep}{-2pt}

\item The Hadoop stack, for better or for worse, has already become the
  {\it de facto} general-purpose, large-scale data processing platform
  of choice. As part of the stack I include higher-level layers such
  as Pig and Hive.

\item Complete, end-to-end, large-data solutions involve heterogeneous
  data sources and must integrate different types of
  processing:\ relational processing, graph analysis, text mining,
  machine learning, etc.

\item No single programming model or framework can excel at every
  problem; there are always tradeoffs between simplicity,
  expressivity, fault tolerance, performance, etc.

\end{list}

\noindent Given these assumptions, the decision criterion I propose is
this:\ in the context of an end-to-end solution, would it make sense
to adopt framework $X$ (HaLoop, Twister, PrIter, Spark, etc.)\ over
the Hadoop stack for solving the problem at hand?\footnote{\small
  Hadoop is already a proven production system, whereas all the
  alternatives are at best research prototypes; let's even say for the
  sake of argument that $X$ has already been made production ready.}
Put another way:\ are the gains gotten from using $X$ worth the
integration costs incurred in building the end-to-end solution? If no,
then operationally, we can consider the Hadoop stack (including
Pig, Hive, etc., and by extension, MapReduce) to be ``good enough''.

Note that this way of thinking takes a broader view of end-to-end
system design and evaluates alternatives in a global
context. Considered in isolation, it naturally makes sense to choose
the best tool for the job, but this neglects the fact that there are
substantial costs in knitting together a patchwork of different
frameworks, programming models, etc. The alternative is to use a
common computing platform that's already widely adopted (in this case,
Hadoop), even if it isn't a perfect fit for some of the problems.

I propose this decision criterion because it tries to bridge the big
gap between ``solving'' a problem (in a research paper) and deploying
the solution in production (which has been brought into stark relief
for me personally based on my experiences at Twitter). For something
to ``work'' in production, the solution must be continuously running;
processes need to be monitored; someone needs to be alerted when the
system breaks; etc. Introducing a new programming model, framework,
etc.\ significantly complicates this process---even mundane things
like getting the data imported into the right format and results
exported to the right location become non-trivial if it's part of a
long chain of dependencies.

A natural counter-argument would be:\ Why should academics be
concerned with these (mere) ``production issues''? This ultimately
comes down to what one's criteria for success are. For me personally,
the greatest reward comes from seeing my algorithms and code ``in the
wild'':\ whether it's an end-to-end user-facing service that millions
are using on a daily basis or an internal improvement in the stack
that makes engineers and data scientists' lives better. I consider
myself incredibly lucky to have accomplished both during my time at
Twitter. I firmly believe that in order for any work to have
meaningful impact (in the way that I define it, recognizing, of
course, that others are guided by different utility functions), how a
particular solution fits into the broader ecosystem is an important
consideration.\footnote{\small As a side note, unfortunately, the
  faculty promotion and tenure process at most institutions does not
  reward these activities, and in fact, some would argue actively
  disincentivizes these activities since they take time away from
  writing papers and grants.}

Different programming models provide different ways of thinking about
the problem. MapReduce provides ``map'' and ``reduce'', which can be
composed into more complex dataflows (e.g., via Pig). Other
programming models are well-suited to certain types of problems {\it
  precisely} because they provide a different way of thinking about
the problem. For example, Pregel provides a vertex-centered approach
where ``time'' is dictated by the steady advance of the superstep
synchronization barriers. We encounter an impedance mismatch when
trying to connect different frameworks that represent different ways
of thinking.  The advantages of being able to elegantly formulate a
solution in a particular framework must be weighed against the costs
of integrating that framework into an end-to-end solution.

To illustrate, I'll present a hypothetical but concrete
example:\ let's say we wish to run PageRank on the interaction graph
of a social network (i.e., the graph defined by {\it interactions}
between users).  Such a graph is {\it implicit} and needs to be
constructed from behavior logs, which is natural to accomplish in a
dataflow language such as Pig (in fact, Pig was exactly designed for
log mining). Let's do exactly that.

With the interaction graph now materialized, we wish to run
PageRank. Consider two alternatives:\ use Giraph,\footnote{\small
  incubator.apache.org/giraph/} the open-source implementation of
Pregel, or implement PageRank directly in Pig.\footnote{\small
  techblug.wordpress.com/2011/07/29/pagerank-implementation-in-pig/}
The advantage of the first is that the BSP model implemented by
Giraph/Pregel is perfect for PageRank and other iterative graph
algorithms (in fact, that's exactly what Pregel was designed to
do). The downside is lots of extra ``plumbing'': munging Pig output
into a format suitable for Giraph, triggering the Giraph job, waiting
for it to finish, and figuring out what to do with the output (if
another Pig job depends on the results, then we must munge the data
back into a form that Pig can use).\footnote{\small Not to mention all
  the error reporting, alerting, error handling mechanisms that now
  need to work across Pig {\it and} Giraph.} In the second
alternative, we simply write PageRank in Pig, with all the
shortcomings of iterative MapReduce algorithms discussed in this
paper. Each iteration might be slow due to stragglers, needless
shuffling of graph structure, etc., but since we likely have the
PageRank vector from yesterday to start from, the Pig solution would
converge mercifully quickly. And with Pig, all of the additional
``plumbing'' issues go away. Given these alternatives, I believe the
choice of the second is at least justifiable (and arguably,
preferred), and hence, in this particular context, I would argue that
MapReduce is good enough.

In my opinion, the arguments are even stronger for the case of stochastic
gradient descent. Why adopt a separate machine-learning framework
simply for running batch gradient descent when it could be seamlessly
integrated into Pig by using stochastic gradient descent and ensemble
methods~\cite{Lin_Kolcz_SIGMOD2012}? This approach costs nothing in
accuracy, but gains tremendously in terms of performance.  In the
Twitter case study, machine learning is accomplished by {\it just
  another Pig script}, which plugs seamlessly into existing Pig
workflows.

To recap:\ Of course it makes sense to consider the right tool for the
job, but we must also recognize the cost associated with switching
tools---in software engineering terms, the costs of integrating
heterogeneous frameworks into an end-to-end workflow are non-trivial
and should not be ignored. Fortunately, recent developments in the
Hadoop project promise to substantially reduce the costs of integrating
heterogeneous frameworks:\ Hadoop Next\-Gen (aka YARN) introduces a
generic resource scheduling abstraction that allows multiple
application frameworks to co-exist on the same physical cluster. In this context,
MapReduce is just one of many possible application frameworks; others include
Spark\footnote{\small github.com/mesos/spark-yarn} and
MPI.\footnote{\small issues.apache.org/jira/browse/MAPREDUCE-2911}
This ``meta-framework'' could potentially reduce the costs of
supporting heterogeneous programming models---an exciting future
development that might let us have our cake and eat it too.  However,
until YARN proves itself in production environments, it remains an
unrealized potential.

\section{Constructive Suggestions}

Building on the arguments above and reflecting on my experiences over
the past several years working on ``big data'' in both academia and
industry, I'd like to make the following constructive suggestions:

\smallskip \noindent {\bf Continue plucking low hanging fruit}, or,
refine the hammer we already have.  I do not think we have yet
sufficiently pushed the limits of MapReduce in general and the Hadoop
implementation in particular. In my opinion, it may be premature to
declare it obsolete and call for a fresh ground-up
redesign~\cite{Battre_etal_SoCC2010,Borkar_etal_EDBT2012}. MapReduce
is less than ten years old, and Hadoop is even younger. There has
already been plenty of interesting work within the confines of Hadoop,
just from the database perspective:\ integration with a traditional
RDBMS~\cite{Abouzeid_etal_VLDB2009,Bajda-Pawlikowski_etal_SIGMOD2011},
smarter task
scheduling~\cite{Zaharia_etal_OSDI2008,Zaharia_etal_2009}, columnar
layouts~\cite{HeYongqiang_etal_2011,LinYuting_etal_SIGMOD2011,Floratou_etal_VLDB2011,Jindal_etal_SoCC2011,Kaldewey_etal_EDBT2012},
embedded indexes~\cite{Dittrich_etal_VLDB2010,Dittrich_etal_VLDB2012},
cube materialization~\cite{Nandi_etal_ICDE2011}, and a whole cottage
industry on efficient join
algorithms~\cite{Blanas_etal_SIGMOD2010,Okcan_Riedewald_SIGMOD2011,LeeRubao_etal_2011,KimYounghoon_Shim_ICDE2012};
we've even seen ``traditional'' HPC ideas such as work stealing make
its way into the Hadoop context~\cite{Kwon_etal_SIGMOD2012}. Much more
potential remains untapped.

The data management and distributed systems communities have developed
and refined a large ``bag of tricks'' over the past several
decades. Researchers have tried applying many of these in the Hadoop
context (see above), but there are plenty remaining in the bag waiting
to be explored. Many, if not most, of the complaints about Hadoop
lacking basic features or optimization found in other data processing
systems can be attributed to immaturity of the platform, not any
fundamental limitations. More than a ``matter of implementation'',
this work represents worthy research. Hadoop occupies a very different
point in the design space when compared to parallel databases, so the
``standard tricks'' often need to be reconsidered in this new context.

So, in summary, let's fix all the things we have a good idea how to
fix in Hadoop (low-risk research), and then revisit the issue of
whether MapReduce is good enough. I believe this approach of
incrementally refining Hadoop has a greater chance of making impact
(at least by my definition of impact in terms of adoption) than a
strategy that abandons Hadoop. To invoke another clich\'{e}: let's
pluck all the low-hanging fruit first before climbing to the higher
branches.

\smallskip \noindent {\bf Work on game-changers}, or, develop the
jackhammer. To displace (or augment) MapReduce, we should focus on
capabilities that the framework fundamentally cannot support. To me,
faster iterative algorithms, illustrated with PageRank or gradient
descent aren't ``it''---given my above arguments on how
for those, MapReduce is ``good enough''. I propose two potential game
changers that reflect pain points I've encountered during my time in
industry:

First, real-time computation on continuous, large-volume streams of
data is not something that MapReduce is capable
of. MapReduce is fundamentally a batch processing framework---and
despite efforts in implementing ``online''
MapReduce~\cite{Condie_etal_2010}, I believe solving the general
problem requires something that looks very different from the current
architecture. For example, let's say I want to keep track of the top
thousand most-clicked URLs posted on Twitter in the last $n$ minutes.
The current solution is to run batch MapReduce jobs with increasing
frequency (e.g., every five minutes), but there is a fundamental limit
to this approach (job startup time), and (near) real-time results are
not obtainable (for example, if I wanted up-to-date results over the
last 30 seconds).

One sensical approach is to integrate a stream processing
engine---a stream-oriented RDBMS
(e.g.,~\cite{Carney_etal_VLDB2002,Gehrke_2003,KrishnamurthyS_etal_SIGMOD2010}),
S4~\cite{Neumeyer_etal_2010}, or Storm\footnote{\small
  github.com/nathanmarz/storm}---with Hadoop, so that the stream
processing engine handles real-time computations, while Hadoop
performs aggregate ``roll ups''. More work is needed along these
lines, and indeed researchers are already beginning to explore this
general direction~\cite{Chandramouli_etal_ICDE2012}. I believe the
biggest challenge here is to seamlessly and efficiently handle queries
across vastly-different time granularities:\ from ``over the past 30 seconds''
(in real time) to ``over the last month'' (where batch computations with
some lag would be acceptable).

Second, and related to the first, real-time interactions with large
datasets is a capability that is sorely needed, but is something that
MapReduce fundamentally cannot support. The rise of ``big data'' means
that the work of data scientists is increasingly important---after
all, the value of data lie in the insights that they generate for an
organization. Tools available to data scientists today are
primitive:\ Write a Pig script and submit a job. Wait five minutes for
the job to finish. Discover that the output is empty because of the
wrong join key. Fix simple bug. Resubmit. Wait another five
minutes. Rinse, repeat. It's fairly obvious that long debug cycles
hamper rapid iteration.  To the extent that we can provide tools to
allow rich, interactive, incremental interactions with large data
sets, we can boost the productivity of data scientists, thereby
increasing their ability to generate insights for the organization.

\smallskip \noindent {\bf Open source everything.} Open source
releasing of software should be the default for any work that is done
in the ``big data'' space. Even the harshest critic would concede that
open source is a key feature of Hadoop, which facilitates rapid
adoption and diffusion of innovation. The vibrant ecosystem of
software and companies that exist today around Hadoop can be
attributed to its open source license.

Beyond open sourcing, it would be ideal if the results of research papers 
were submitted as patches to existing open source software (i.e.,
associated with JIRA tickets). An example is recent work on distributed cube
materialization~\cite{Nandi_etal_ICDE2011}, which has been submitted
as a patch in Pig.\footnote{\small
  issues.apache.org/jira/browse/PIG-2167} Of course, the costs
associated with this can be substantial, but this represents a great
potential for collaborations between academia and industry; committers
of open source projects (mostly software engineers in industry) can
help shepherd the patch. In many cases, transitioning academic
research projects to production-ready code make well-defined summer
internships at companies. These are win-win scenarios for all:\ the
company benefits immediately from new features; the community benefits
from the open sourcing; and the students gain valuable experience.

\section{Conclusion}

The clich\'{e} is ``if all you have is a hammer, then everything looks
like a nail''. I argue for going one step further:\ ``if all you have
is a hammer, throw away everything that's not a nail''! It'll make
your hammer look amazingly useful. At least for some time. Soon or
later, however, the flaws of the hammer will be exposed---but let's
try to get as much hammering done as we can before then. While
we're hammering, though, nothing should prevent us from developing
jackhammers.

\end{document}